\documentstyle[prl,epsfig,aps]{revtex}
\begin{document}
\preprint{LPTENS-97/18}
\twocolumn[\hsize\textwidth\columnwidth\hsize\csname@twocolumnfalse\endcsname
\title{Quantum aging in mean-field models}
\author{Leticia F. Cugliandolo{$^*$}\cite{add1} and
Gustavo Lozano{$^{**}$}\cite{add2}}
\address{ 
$^{*}$
Laboratoire de Physique Th\'{e}orique de l'Ecole Normale 
Sup\'{e}rieure,
24 rue Lhomond, F-75231 Paris Cedex 05, France \\
and Laboratoire de Physique Th\'eorique  et Hautes Energies \\
Univ. Paris VI and VII, 
5 \`eme \'etage,  Tour 24,
4 Place Jussieu, 75252 Paris Cedex 05, France\\
$^{**}$Division de Physique Th\'eorique, IPN,
Universit\'e de Paris XI, F-91400, Orsay, France
}
%\date\today
\maketitle
\begin{abstract}
We study the real-time dynamics of quantum 
models with long-range interactions coupled to a heat-bath within the 
closed-time path-integral formalism. We
show that quantum fluctuations depress the transition temperature.
In the subcritical region
there are two asymptotic time-regimes with (i) stationary, 
and (ii) slow aging dynamics.
%with a waiting-time dependent 
%decoherce time.
We extend the quantum fluctuation-dissipation theorem to the
nonequilibrium case in a consistent way with the notion of an effective
temperature that drives the system in the aging regime. 
The classical results are recovered for $\hbar\to 0$.       
$\;$ \newline
$\;$ \newline

\end{abstract}

\twocolumn 
\vskip.5pc]
\narrowtext
The dynamics of nonequilibrium 
systems is being intensively studied now. 
Notably, glassy systems below their 
critical temperature have a very slow evolution with nonstationary
dynamics\cite{St}. Several theoretical ideas\cite{review} are used to 
describe it, namely, scaling arguments, phase space models,
analytical solutions to mean-field models
and numerical simulations.
The analysis of simple mean-field models (with long-range interactions) 
has provided a general scenario\cite{Cuku,Cuku2} 
that is now being verified numerically for 
more realistic models\cite{numerics,numerics2}. 
All these studies concern classical systems.

Recent experiments \cite{qexp} have motivated a renewed interest on the 
effect of quantum fluctuations (QF) on glassy systems.
Up to present, theoretical  studies
have focused on  how QF affect their {\it equilibrium} 
properties\cite{quantum,quantum2,quantumreplicas,Giledou}. 

Since glasses below $T_g$ are not expected to 
reach equilibrium in experimentally accessible times, it is important
to device a method to understand the 
influence of QF on the trully nonequilibrium 
{\it real-time} dynamics of this type of systems. 
Intuitively, 
one expects QF to only affect
the short-time dynamics; however, they are also expected to act as 
thermal fluctuations. It is then not clear whether QF would 
destroy glassiness or modify it drastically.  

Our aims are 
(i) to present a formalism suited to 
study the real-time dynamics of a nonlinear, possibly disordered, model
in contact with a bath;
(ii) to propose a framework to study its dynamics
that could be also 
applicable to more realistic, finite dimensional, models;
(iii) to show that below a critical line 
QF do not destroy the nonequilibrium effects
of the glassy phase and that QF add up to
an effective temperature $T_{\sc eff}$;
(iv) to prove that $T_{\sc eff}$ is non-zero even at 
zero bath temperature, that it
drives the dynamics at late epochs and that it makes the dynamics 
appear classical in that time-regime.      
A longer account of our results will appear 
elsewhere.
 
The real-time dynamics of a quantum system 
is described with a closed-time path-integral
generating functional (CTP-GF) \cite{closedpath}. 
%A doubling of degrees of freedom produced by the 
%introduction of two sources, needed to 
%compute time-ordered correlation functions.
We choose a set of noninteracting harmonic oscillators with an 
adequate distribution of frequencies as a bath,
and a linear interaction between  
bath and system\cite{Feve}. 
The bath variables are next integrated out and 
the effect of the bath manifiests 
in the effective action through two non-local kernels associated to  
dissipation ($\eta$) and noise ($\nu$).
If the model is disordered, one needs to compute averaged expectation values. 
However,
the CTP-GF without sources is independent
of the realization of disorder 
and one can hence avoid, in a quantum dynamical calculation, 
the introduction of replicas. In addition, when $\hbar\to 0 $, the CTP-GF
yields the classical Martin-Siggia-Rose one.

For the sake of concreteness, we study the $p$ spin-glass
\begin{equation}
H_J[{\bbox \phi}]  
= 
\frac{1}{2 m N} \sum_{i=1}^N \Pi_i^2 +
\sum_{i_1,\dots,i_p} J_{i_1 \dots i_p} \phi_{i_1} 
\dots \phi_{i_p}  
\end{equation}
with $\Pi_i$ the canonical momenta, $[\Pi_i,\phi_j]=-i\hbar \delta_{ij}$.
The multispin interactions $J_{i_1\dots i_p}$ are taken from a Gaussian
distribution with zero mean and variance $\tilde J^2 p!/(2 N^{p-1})$ and
$\sum_{i=1}^N [\langle \phi_i^2(t) \rangle]_J =N$, 
$\forall t$. Square brackets denote an average over disorder and
$\langle \bullet \rangle$ the average over temporal histories.
The quantum mean-field equations 
follow from a saddle-point approximation and
involve
the symmetrized auto-correlation 
$N C(t,t_w) \equiv 
[\langle {\bbox \phi}(t) {\bbox \phi}(t_w) + 
{\bbox \phi}(t_w) {\bbox \phi}(t) \rangle]_J$
and
the response to an infinitesimal perturbation $h$ applied at time $t_w$,
$N R(t,t_w) \equiv \delta [ \langle {\bbox \phi}(t) \rangle 
/\delta {\bbox h}(t_w) ]_J|_{h=0}$.
We define 
the vertex and self-energy:
\begin{eqnarray*}
D(t,t_w)&=&\tilde D(t,t_w) + 2 \hbar \nu(t-t_w)
\nonumber\\ 
\Sigma(t,t_w) &=&\tilde \Sigma(t,t_w) - 4 \eta(t-t_w)
\end{eqnarray*}
respectively.
$\tilde \Sigma$ and $\tilde D$ are functions of $C$ and $R$  
that follow either from the average over disorder or 
from approximations of  
the nonlinear interactions in the model 
(e.g. mode-couling approximations). For the $p$-spin model 
and $t\geq t_w$ they read
\begin{eqnarray*}
\tilde D(t,t_w) + \frac{i\hbar}{2} \Sigma(t,t_w)
=
\frac{p\tilde J^2}{2} \left( 
C(t,t_w) + \frac{i\hbar}{2} R(t,t_w)
\right)^{p-1}
\end{eqnarray*}   
The terms associated to the bath are
\begin{eqnarray}
\nu(t-t_w) &=&
\int_0^\infty d\omega I(\omega) \coth(\beta\hbar\omega/2) \cos(\omega(t-t_w))
\; ,
\nonumber\\
\eta(t-t_w) &=& \theta(t-t_w)
\int_0^\infty d\omega I(\omega) \sin(\omega(t-t_w))
\; .
\nonumber
\end{eqnarray}
We chose an Ohmic distribution of oscillator frequencies 
$
I(\omega) =  (M \gamma_o/\pi) \;  \omega \; \exp(-|\omega|/\Lambda)
$
with $\Lambda$ a cut-off and $M \gamma_o$ a constant 
that plays the r\^ole of a friction coefficient.  
Their particular form depends on the choice of bath. 
The equations, for a random initial condition, are
\begin{eqnarray}
R(t,t_w) &=& G_o(t,t_w) +\int_0^t \int_0^t du dv 
G_o(t,u) \Sigma(u,v) R(v,t_w)
\nonumber\\
C(t,t_w) &=& \int_0^t du \int_0^{t_w} dv 
R(t,u) D(u,v) R(t_w,v)
\label{dynamics}
\end{eqnarray}
with $G_o^{-1}(t,u)=\delta(t-u) \left( m\partial_u^2+\mu(t) \right)$ 
the propagator and $\mu(t)$ either a Lagrange multiplier enforcing a spherical
constraint or the strength of a  quadratic term. 
In the first case it is determined by the `gap' equation:
$
\mu(t) = \int_0^t du \, 
\Sigma(t,u) C(t,u) + D(t,u) R(t,u) - m \left. \partial_{t^2} C(t,t_w)
\right|_{t_w\to t^{-}}
$.
Causality implies $R(t,t_w) = \Sigma(t,t_w)=0$ if $t_w > t$.
The inertial term imposes continuity on the equal-times correlator 
$\lim_{t_w\to t^-} \partial_t C(t,t_w) =
\lim_{t_w\to t^+} \partial_t C(t,t_w) = 0$, 
and $R(t,t)=0$, $\lim_{t_w\to t} \partial_t R(t,t_w)=1/m$.
In what follows we set $t \geq t_w$. 
The coupling to the bath implies dissipation; if $M\gamma_o\neq 0$
the energy density ${\cal E}$ 
of the system decreases. One can then envisage to 
switch-off the coupling (set $M\gamma_o=0$) when ${\cal E}$ reaches
a desired value and follow the subsequent evolution at constant ${\cal E}$.  
This would be useful to further understand the energy landscape. 
Here, we keep $M \gamma_o\neq 0$ for all times, 
reparametrize time 
according to $t \to M \gamma_o t$ and 
transform $\hbar$ in a free parameter $\hbar \to M\gamma_o \hbar$.
Consistently, 
$C \to C$, $R \to R/(M\gamma_o)$, $m \to (M\gamma_o)^2 m$, 
$\tilde J \to \tilde J$, $\beta \to \beta$,$\Lambda\to\Lambda/(M\gamma_o)$
and the units are
$[C]=[R]=[\hbar]=1$, $[m]=[\beta]=[1/\tilde J]=[1/\Lambda]=[t]$.

We focus on $p>2$, since $p=2$ needs a special treatment\cite{cosmology}. 
Since Eqs. (\ref{dynamics}) are causal, we  
contruct its solution numerically, step by step in $t$, 
spanning $0 \leq t_w \leq t$. 
The numerical and analytical studies yield:

(i)  
{\it Quantum mode-coupling equations}.
For $\hbar$ and $T$ above a critical line, there is a finite 
equilibration time $\tau_{\sc eq}$ after which equilibrium 
dynamics sets in. The solution satifies 
invariance under time-translations (TTI) 
and the quantum fluctuation-dissipation theorem (QFDT). 
It is a `paramagnetic/liquid'
phase. A  TTI-QFDT  ansatz 
yields
\begin{eqnarray}
R(\omega) &=& \frac{1}{-m \omega^2 + \mu_\infty - \Sigma(\omega)}
\; ,
\label{111}
\\
C(\omega) &=& D(\omega) |R(\omega)|^2
\label{222}
\; .
\end{eqnarray}
with the couples $R$ and $C$, and $\Sigma$ and $D$,
verifying QFDT:
\begin{eqnarray}
& & 
%R(\omega) = - \frac{2}{\hbar} \lim_{\epsilon\to  0^+} 
%\int \frac{d\omega'}{2\pi} \; 
%\frac{\tanh\left( (\beta\hbar\omega')/2\right)}
%{\omega-\omega'+i \epsilon} \,
% \, C(\omega)
R(\tau) = 
\frac{2i}{\hbar} \theta(\tau)
\int \frac{d\omega}{2\pi} \, 
e^{-i\omega\tau}
\tanh\left( \frac{\beta\hbar\omega}{2}\right)
C(\omega)
\label{FDT1}
%\; 
%\\
%& & 
%C(z) + \frac{i\hbar}{2} R(z) 
%=
%C(z^*) - \frac{i\hbar}{2} R(z^*)
%\; ,
%\label{FDT2}
\end{eqnarray}
($\tau=t-t_w$). Equations (\ref{111})-(\ref{222}) 
are the quantum version of the mode-coupling equations 
used to describe supercooled-liquids\cite{Go}.  
Away from the critical line, $C$ and $R$ decay to zero 
very fast, with osicllations. 
Approaching the critical line $T_d(\hbar_d)$, the decay 
slows down and, if $T_d(\hbar_d)\neq 0$, a 
plateau develops in $C$. At the critical line, the 
lenght of the plateau tends to infinity. The quantum critical point
$(T_d=0,\hbar_d\neq 0)$ is discussed below.

(ii) {\it Dynamics in the glassy phase.} 
For  $\hbar$ and $T$ below the critical line, 
$\tau_{\sc eq} \to \infty$ (as a function of 
 $N$):
times are always finite with respect to $\tau_{\sc eq}$.
The system does not reach equilibrium. 
This is a `spin-glass/glass' phase.
There are two time regimes
with different behaviours according to the relative value of $t-t_w$ and 
a characteristic (model-dependent) time ${\cal T}(t_w)$:

If $t-t_w \leq {\cal T}(t_w)$ ($C(t,t_w) \geq q$) 
the dynamics is {\it stationary}; TTI and QFDT hold. 
In other words,  
\begin{equation}
q + C_{\sc st}(t-t_w) = \lim_{t_w\to\infty} C(t,t_w)
\end{equation}
and $R_{\sc st}(t-t_w)=\lim_{t_w\to\infty} R(t,t_w)$.
The correlation approaches a plateau $q$ since $\lim_{t-t_w\to\infty} C_{\sc st}(t-t_w)=0$.
The response  approaches zero, $\lim_{t-t_w\to\infty} R_{\sc st}(t-t_w)=0$.
The equations for $C_{\sc st}$ and $R_{\sc st}$ are identical 
to Eqs.(\ref{111})-(\ref{222}) apart from contributions to 
$\mu_\infty$ coming from the aging regime. 
$C_{\sc st}$ and $R_{\sc st}$ are linked through Eq.(\ref{FDT1}).

 If $t-t_w > {\cal T}(t_w)$  ($C(t,t_w) < q$) 
the dynamics is {\it nonstationary}; TTI and FDT do not hold and 
there is quantum aging. The correlation decays from $q$ to $0$ and 
we call it $C_{\sc ag}(t,t_w)$. 
The decay of $C$ becomes {\it monotonous} in the aging regime; 
the properties of monotonous two-time correlation functions derived in 
Ref.\cite{Cuku2} imply:
\begin{equation}
C_{\sc ag}(t,t_w) = \jmath^{-1}\left( \frac{h(t_w)}{h(t)} \right)
\label{jmath}
\; ,
\end{equation} 
with $h(t)$ a growing function of time. For the $p$-spin model,
$\jmath^{-1}$ is the identity. 

The system has a weak long-term memory, 
meaning that the response tends to  zero {\it but} 
its integral over a growing time-interval is {\it finite}.
The QFDT is more involved than the classical one 
since it is non-local (Eq.(\ref{FDT1})). 
In the nonequilibrium case, we  
generalize it to 
\begin{eqnarray}
R(t,t_w) &=& \frac{2 i}{\hbar} \; \theta(t-t_w)
\int_{-\infty}^\infty \frac{d\omega}{2 \pi}  \; 
\exp(-i \omega(t-t_w)) 
\nonumber \\
& & 
\times \tanh\left( \frac{X(t,t_w) 
\beta \hbar \omega}{2} \right) \, C(t,\omega)
\label{XFDT1}
\\
C(t,\omega)
&\equiv& 
2 {\mbox{Re}} \int_0^t ds \exp(i \omega (t-s)) C(t,s)
\label{XFDT2}
\; 
\end{eqnarray}
with $X(t,t_w)$ a function of $t$ {\it and} $t_w$.  
If the evolution is TTI 
and $X(t,t_w)=1$ for all times, Eq.(\ref{XFDT1}) 
reduces to Eq.(\ref{FDT1}). 
Interestingly enough, $T_{\sc eff} \equiv T/X(t,t_w)$ acts as 
an effective temperature in the system\cite{Cukupe}. 
For a model with two two-time sectors we propose
\begin{eqnarray}
X(t,t_w)&=&
\left\{
\begin{array}{ll}
X_{\sc st} =1
\;\;\; &{\mbox{if}} \;\;\;
t-t_w \leq {\cal T}(t_w) 
\nonumber\\
X_{\sc ag}(\hbar,T)
\;\;\; &{\mbox{if}} \;\;\; 
t-t_w > {\cal T}(t_w)
\end{array}
\right.
\; .
\end{eqnarray}
with $X_{\sc ag}$ a non-trivial function of
$\hbar$ and $T$. 
When $t$ and $t_w$ are widely separated, the integration over $\omega$ in 
Eq.(\ref{XFDT1}) is dominated by $\omega\sim 0$. Therefore, the factor 
$\tanh(X_{\sc ag}(t,t_w) \beta\hbar\omega/2)$ can be substituted by 
$X_{\sc ag}\beta\hbar\omega/2$ 
({\it even at} $T=0$ if $X_{\sc ag}(\hbar,T) =x(\hbar) T$ when $T\sim 0$). 
Hence, 
\begin{equation}
R_{\sc ag}(t,t_w) \sim \theta(t-t_w) X_{\sc ag} \beta
\partial_{t_w} C_{\sc ag}(t,t_w)
\label{Xclassical}
\end{equation}
and one recovers, in the aging regime, the {\it classical} modified 
FDT\cite{Cuku,Giledou}. 
Numerically (and experimentally), 
it is simpler to check it  using the 
integrated response $\chi(t,t_w)=\int_{t_w}^t dt'' R(t,t'')$ over a large time
window\cite{Cuku2,numerics2}, that reads
\begin{eqnarray}
\chi(t,t_w)&=& 
\left\{
\begin{array}{l}
\int_0^{t-t_w} d\tau' R_{\sc st}(\tau') 
\nonumber\\
\int_0^\infty d\tau' R_{\sc st}(\tau') +
\frac{X_{\sc ag}}{T} (q-C_{\sc ag}(t,t_w)) 
\; 
\end{array}
\right.
\end{eqnarray}
the first line holds for $C(t,t_w) > q$ while the second 
holds for $C(t,t_w)<q$.
Remarkably, the correlation {\it and} 
the violation of QFDT are given by similar expressions to the 
classical -- though the values of $q$ and $X_{\sc ag}$
depend on $\hbar$. In this sense, we say that ${\cal T}(t_w)$ acts a 
time-dependent `decoherence time', beyond which the nonequilibrium
regime is `classical'.

\begin{figure}
\centerline{\epsfxsize=9cm
%\epsffile{ClogL5hb01T0.ps}
\epsffile{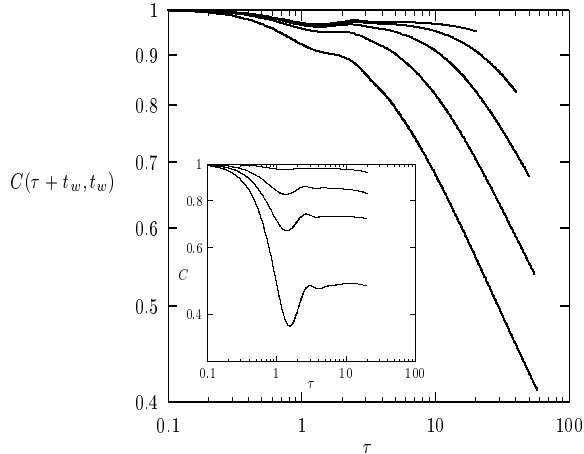}
}
\caption{The correlation function $C(\tau+t_w,t_w)$ vs $\tau$ 
for the $p=3$ SG model, $\Lambda=5$, $\tilde J=1$, $m=1$,  
$T=0$ and $\hbar=0.1$. The waiting times are, 
from bottom to top, 
$t_w=2.5,5,10,20,40$. $q\sim 0.97$. In the inset, 
the same curves for $t_w=40$ and, from top to bottom, $\hbar=0.1,0.5,1,2$.
}
\label{figcorr}
\end{figure}

An ansatz like Eq.(\ref{jmath}) for $C_{\sc ag}$ combined with 
Eq.(\ref{Xclassical})  for the FDT violation 
solves the equations in the aging regime. It yields
\begin{eqnarray}
1 &=& \left(R_{\sc st}(\omega=0) \right)^2
{\tilde J}^2 p(p-1)q^{p-2} /2 
\label{qEA}
\\
X_{\sc ag}/T 
&=& 
%\sqrt{ \frac{2 (p-2)^2}{p(p-1) q^p} }
R_{\sc st}(\omega=0) (p-2)/q 
\label{X}
\end{eqnarray} 
that become  ${\tilde J}^2 p(p-1)/2 q^{p-2} (1-q)^2
$ $=$ $2T^2$ and 
$X_{\sc ag}$ $=$ $(p-2)(1-q)/q$ when  $\hbar\to 0$ \cite{Cuku}.

QF depress the critical temperature. The transition line $T_d(\hbar_d)$
ends at a quantum critical point $(T_d=0,\hbar_d\neq 0)$. Equations 
(\ref{qEA})-(\ref{X}) indicate that when the transition occurs at
$T_d(\hbar_d)\neq 0$ it is first-order with
$X_{\sc ag}\to 1$, $q\to q_d\neq 0$ and a finite linear
stationary susceptibility $R_{\sc st}(\omega=0)<+\infty$
(as in the classical limit). On this line, $q_d \propto T_d^{2/p}$
and $q_d\to 0$ for $T_d\to 0$.
At the quantum critical point,
Eqs.(\ref{qEA})-(\ref{X}) suggest a second-order transition;
if $q_d$ tends to zero as $q_d \sim (\hbar_d-\hbar)^{\alpha p/2}$,
then $X_{\sc ag}/T \sim (\hbar_d-\hbar)^{-\alpha}$, 
and $R_{\sc st}(\omega=0)\sim (\hbar_d-\hbar)^{\alpha(1-p/2)}$
diverges when $p > 2$ ($\alpha$ is positive).

Let us now discuss the numerical results.
In all figures $p=3$,  
$\Lambda=5$, $\tilde J=1$ and $m=1$. 
We use $T=0$ and $\hbar=0.1$ to illustrate the dynamics in the glassy
phase. We also discuss the dependence upon $T$ and $\hbar$.   

\begin{figure}
\centerline{\epsfxsize=9cm
\epsffile{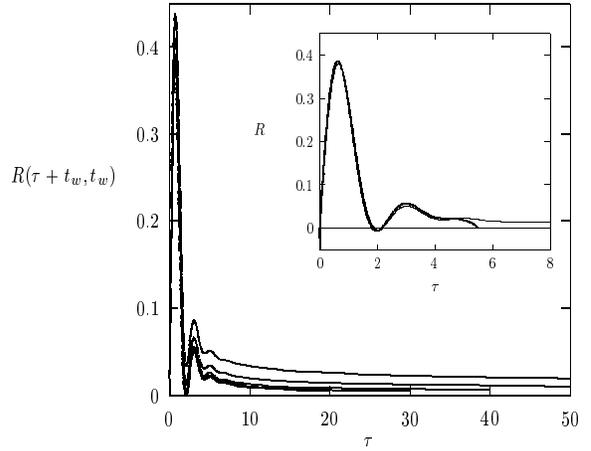}
}
\caption{The response function for the same model as above. 
The waiting-times increase from top to bottom.
In the inset, check of FDT in the stationary regime.  
The full line is $R(t,t_w)$ for $t=40$ fixed and $t_w\in [0,40]$.
The dots are obtained from Eq.(\ref{XFDT1}) with $X_{\sc st}=1$, 
using the numerical data for $C_{\sc st}(t-t_w)=C(t,t_w)-q$ 
($q\sim 0.97$, see Fig.\ref{figcorr}). 
In both cases 
the response is plotted against $t-t_w$.
}
\label{figresp}
\end{figure}

In Figs. \ref{figcorr} and 
\ref{figresp} we show the correlation 
$C(\tau+t_w,t_w)$ (log-log plot) and the response $R(\tau+t_w,t_w)$ (linear plot) 
vs the subsequent time 
$\tau$. 
These plots  demonstrate the existence of the stationary and aging regimes. 
For $t-t_w < {\cal T}(t_w)$ (e.g. 
${\cal T}(40) \sim 5$) TTI  and FDT are established while
beyond ${\cal T}(t_w)$ they break down.
For $\hbar=0.1$ the plateau in $C$ is at $q\sim 0.97$. 
$C$ oscillates around $q$ but 
is monotonous when it goes below it. 
In the inset we present the dependence of $q$ on $\hbar$ for $T=0$.
QF generate a $q<1$ such that the larger $\hbar$ the smaller
$q$. The addition of thermal fluctuations has a similar effect, 
the larger $T$, the smaller $q$.
In order to check FDT in the stationary regime,
in the inset of Fig. \ref{figresp} 
we compare $R(t,t_w)$ from the numerical 
algorithm for $t=40$ fixed and $t_w\in[0,40]$
(full line)
with $R(t,t_w)$ from Eq.(\ref{XFDT1}) with $X=1$ 
using $C_{\sc st}(t-t_w)=C(t,t_w)-q$, $q\sim 0.97$ obtained from 
the algorithm (dots).
The accord is very good if $t-t_w \leq {\cal T}(t_w) \sim 5$.   

\begin{figure}
\centerline{\epsfxsize=9cm
\epsffile{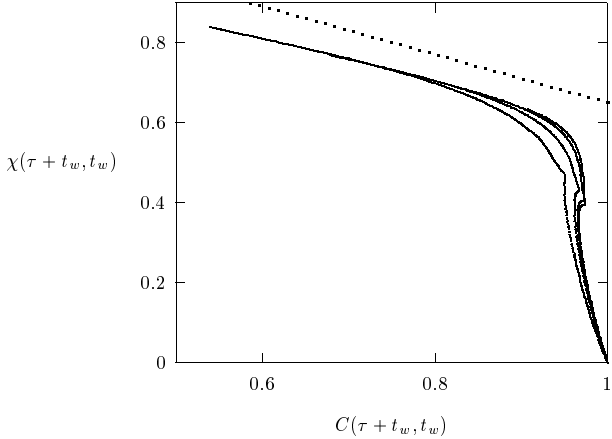}
}
\vspace{.1cm}
\caption{The integrated response $\chi(\tau+t_w,t_w)$ vs. the symmetrized 
auto-correlation function $C(\tau+t_w,t_w)$ in a parametric plot. The 
curves correspond to $t_w=10,20,30,40$. $T=0$ and $\hbar=0.1$. The 
dots are a guide to the eye and 
represent the analytic result $X_{\sc ag}/T \sim 0.60$ for the slope of the
asymptotic curve in the aging regime $C<q\sim 0.97$.  
}
\label{figchiC}
\end{figure}

In Fig.\ref{figchiC} we plot the integrated response $\chi$ 
vs. $C$ in a parametric plot. When $C < q \sim 0.97$ the 
$\chi$ vs $C$ curve approaches a straight line of {\it finite} 
slope $1/T_{\sc eff} = X_{\sc ag}/T$ that we 
estimate from Eqs.(\ref{qEA}) and (\ref{X}) as  $X_{\sc ag}/T \sim 0.60$. 
The dots are a guide to the eye representing it. 

To summarize, this formalism provides a clear framework to study
the nonequilibrium regime and/or 
the eventual approach to equilibrium of quantum
systems. It circumvents the difficulties of performing 
numerical simulations in real-time and predicts the existence 
and the structure of a rich nonequilibrium regime for glassy 
systems even in the presence of QF. Based on the success of 
mean-field like classical glassy models to describe  some 
aspects of the  dynamics of realistic glassy  
systems (see \cite{review,Go} and references therein)
one can expect that quantum mean-field models of the type considered 
here do also capture important features of the real-time dynamics 
of real systems.

\vspace{.2cm}
\noindent
ACKNOWLEDGEMENTS. We wish to thank T. Evans, 
A. Georges, D. Grempel, T. Giamarchi, J. Kurchan, P. Le Doussal,
R. Revers 
and M. Rozenberg for useful 
discussions. G.L is supported by the EC grant N ERBFMBICT 961226.

\end{document}